# A Discussion of Special Relativity

Galina Weinstein[*]

Five topics: A rigid body does not exist in the special theory of relativity; distant simultaneity defined with respect to a given frame of reference without any reference to synchronized clocks; challenges on Einstein's connection of synchronization and contraction; a theory of relativity without light, composition of relative velocities and space of relative velocities.

## 1. Kinematics of the "Rigid Body" – No such Thing

In 1905, Einstein wrote in his relativity paper, "On the Electrodynamics of Moving Bodies": "The theory to be developed here is based, like all electrodynamics, on the kinematics of the rigid body, since the assertions of any such theory concerns with the relations among rigid bodies (coordinate systems), clocks, and electromagnetic processes".[1] Einstein defined position by "means of rigid measuring rods and using the methods of Euclidean geometry".[2]

John Stachel explains that according to special relativity information cannot travel faster than the speed of light. Thus there can be no *rigid body*, which is possible in classical mechanics where forces are transferred at infinite speeds. A rigid body moves in a rigid manner, no matter what forces are imposed on the body. In fact, *rigid motions* can be defined without any contradiction in special relativity, even though a rigid body does not exist in the special theory of relativity.

Einstein spoke about rigid body because in 1905 he did not realize that this concept of the rigid body is incompatible with the special theory of relativity, and must be replaced by the concept of rigid motions.

At the 81st meeting of the Gesellschaft Deutscher Naturforscher und Ärzte in Salzburg in September 1909, Max Born presented a paper in which he first analyzed this problem and showed the existence of a very limited class of rigid motions in special relativity; although one cannot have rigid bodies in special relativity, one can move the measuring rod rigidly, and Born gave a Lorentz invariant definition for that.[3]

[*] Written while at the Center for Einstein Studies, Boston University

Born explained that, "measuring rods that maintain their length at uniform translation in the co-moving coordinate system, suffer a contraction in the direction of their velocity when viewed from the system at rest. Therefore, the concept of the rigid body drops out [fällt], at least in its form adapted to *Newton*ian kinematics".[4]

Sommerfeld had commented on Born's talk in a "Diskussion"[5] following the paper. Born noted in a latter paper[6] that, he had discussed with Einstein the subject and they were puzzled that the rigid body at rest can never be brought into uniform motion.

In 1911 in his paper that discusses the problem of rigid bodies in the theory of relativity, "Zur Diskussion über den starren Körporen in der Relativitätstheorie", Max Laue responded to Paul Ehrenfest's paradox and to Born's paper. He wrote that the necessity to admit the possibility of shape changes at any body is also explained in a useful way by the example given by Einstein in the conversation (with Born), which was communicated to him in the discussion at the meeting of natural scientists in Königsberg by Sommerfeld.[7]

A measuring rod is at rest, while another one (forming an angle $\alpha$ with the first) moves with velocity $q$ perpendicular to its direction. The intersection point of both measuring rods travels over the first rod with velocity, $V = q/sin\alpha$, and if $q = 1$ cm/sec, then $V$ becomes greater than the speed of light $c$ by the choice of sufficiently small values for $\alpha$. This is a purely geometric conclusion which cannot be changed by any physical theory. Laue asked, whether this fact contradicts Einstein's theorem according to which, physical effects cannot propagate with superluminal speed $V > 1$?[8]

Laue then explained that the measuring rod is again now at rest. At the beginning the second measuring rod is at rest. At one end $A$ of the first rod – the intersection point between the two rods – an event happens. The second rod is immediately set into motion with velocity $q$ (previously defined). Then the other end $B$ would have to go into motion immediately or simultaneously as a result of the event that happened at $A$. Laue said that the transmission from $A$ to $B$ takes place with superluminal speed $V$. But special relativity says that no signal can be transferred beyond the fundamental velocity. Therefore the concept of a rigid body contradicts this consequence of special relativity.[9]

According to special relativity the second measuring rod, which is brought into motion by an impulse emanating from $A$, doesn't remain straight but gets shortened in $A$ (i.e., it does not remain constant in length). The previous conclusion, $V > c$, only holds for motions at which the measuring rod remains straight, i.e. when the measuring rod is a rigid body.

In June 9[th], 1952 Einstein wrote an appendix to the fifteenth edition of his popular 1916/1917 book *Über die spezielle und die allgemeine Relativitätstheorie*

*Gemeinverständlich* (*On the Special and the General Theory of Relativity*). In this appendix he wrote,[10]

"The subtlety of the concept of space was enhanced by the discovery that there exist no completely rigid bodies. All bodies are elastically deformable and alter in volume with change in temperature".

## 2. Definition of Distant Simultaneity

In 1905 Einstein had given meaning to the space coordinates by the system of rigid rods at rest in a coordinate system (in a state of inertial motion), and now the question is how to give meaning to the time coordinate. We need to give a definition of simultaneity, because all judgments in which time is involved are always judgments of simultaneous events.[11] Einstein defined measurement conventions using clocks and measuring rods and adopted a definition of distant simultaneity.

### 2.1 1905: Definition of Distant Simultaneity With Reference to Synchronized Clocks

Let us first see how Einstein proceeded with the aid of rods and synchronized clocks to define distant simultaneity. In section §1 Einstein says, "As we know from experience". Einstein implicitly tells his reader: you have been on trains, looked at the hands of clocks: "If for instance, I say, 'That train arrives here at 7 o'clock', I mean something like this: 'This pointing of the small hand of my watch to 7 and the arrival of the train are simultaneous events".[12] Einstein substitutes "the position of the small hand of my watch" for "time". This is reasonable when we want to define time for the place where the watch is located. Einstein uses the train and clock imaginative example to exemplify to his reader some experience from his daily life.

However, Stachel says that we must not infer from such examples that this and other similar thought experiments could indicate that Einstein might have been inspired by patents of clocks, trains and clock towers that happened to stand near the patent office he used to work at in Bern.[13]

Next Einstein tells his reader that if he wants to evaluate time of events *that are remote from the clock*, it could be done, but it is not most convenient. This sounds reasonable as well, but it is a leap of thought.

However, immediately after this intuitive presentation, Einstein says: "We might, of course, content ourselves with evaluating the time of the events determined by an observer stationed together with the clock at the origin of the coordinates, and assigning the corresponding positions of the hands to light signals, given out by every event to be timed, and reaching him through empty space. However, as we know from experience, this coordination has the disadvantage of not being independent of the position of the observer equipped with the clock. We arrive at a much more practical arrangement by the following consideration".[14] Einstein introduces a synchronization procedure:

"If there is a clock at the point A in space, then an observer at A can determine the time of events in the immediate vicinity of A by finding the positions of the hands of the clock, which are simultaneous with these events. If there is at the point B of space another clock – and we should add, 'a clock in all respects resembling the one at A' – it is possible for an observer at B to determine the time values of events in the immediate vicinity of B. But it is not possible, without further assumption, to compare, the time of an event at A with the one at B. We have so far defined only an "A time" and a "B time," but not defined a common "time" for A and B. The latter time can now be defined by establishing *by definition* distant simultaneity that, the "time" required by light to travel from A to B equals the "time" it requires to travel from B to A. Let a ray of light go from A to B at "A time" $t_A$, at "B time" $t_B$ is reflected from B towards A, and arrives again at A at "A time" $t'_A$. By definition the two clocks are synchronous if

$$t_B - t_A = t'_A - t_B. \text{"}[15]$$

Or better,

$$t_A \equiv \frac{t_A - t'_A}{2}.$$

Einstein then assumes two additional hypotheses that are generally valid:[16]

1) Transitivity: if the clock at B runs synchronously with the clock at A, the clock at A runs synchronously with the clock at B.

2) Additivity: if the clock at A runs synchronously with the clock at B as with the clock at C, then the clocks at B and C also run synchronously relative to each other.

Subsequently Einstein says, "With the help of certain (imagined) physical experiments, we have defined what is to be understood by synchronous clocks at rest relative to each other located at different places, and obviously obtained a definition of 'simultaneous' and 'time' ".[17] And also the definition:[18] "The 'time' of an event is the reading that is given simultaneously with the event by a clock at rest, which is located at the place of the event, this clock for all times is synchronous with a specified clock at rest".

Einstein referred again to experience: "In agreement with experience", we further stipulate the quantity:

$$2AB/(t'_A - t_A) = c,$$

be a universal constant (the velocity of light in empty space). Einstein defined time by means of clocks at rest in the system at rest, and he called it "the time of the system at rest".

## 2.2 Definition of Distant Simultaneity without Reference to Synchronized Clocks

In his popular 1916/1917 book, *On the Special and the General Theory of Relativity* (*Über die spezielle und die allgemeine Relativitätstheorie Gemeinverständlich*) Einstein defined distant simultaneity with respect to a given frame of reference without any reference to synchronized clocks.

Five years earlier Varičak criticized Einstein (as discussed further below) and Einstein answered him by suggesting a thought experiment that does not require the use of synchronized clocks in order to verify the Lorentz contraction.

Einstein considers two bolts of lightning that strike points A and B respectively. The interval AB is measured, and an observer is placed in the middle M of the interval AB. The observer is provided with a device (for example two mirrors inclined at an angle of $90^0$ to each other) that allows him simultaneous view of both places A and B. If he perceives the two strokes of lightening simultaneously, then they are simultaneous. Einstein first says that the two optical signals from A to M and B to M both proceed at the same speed. However, the speed could only be defined if we had already at our disposal the means of measuring time. Einstein then adopts a definition of simultaneity, "That light requires the same time to traverse the path A arrow M as for the path B arrow M". Only after defining distant simultaneity does Einstein proceed on to the definition of "time", which does involve clocks.[19]

## 3. Relativity of Simultaneity

## 3.1. 1905: On the Relativity of Lengths and Times

In order to evaluate remote events Einstein needed to formulate the two heuristic principles of his theory; these would guide him in his search for the solution to the problem:[20]

"1. The laws by which the states of physical systems change are independent, whether these changes of state are referred to the one or the other of two systems of coordinates in uniform motion of translation.

2. Any ray of light moves in the coordinate system "at rest" with the definite velocity c, independent of whether the ray is emitted by a body at rest or in motion. Here:

Velocity = light path/time interval".

Where "time interval" is understood in the sense of the definition of distant simultaneity given in §1 for clock synchronization in one system: the time required by light to travel from A to B = time it requires to travel from B to A.

With the aid of the two principles and the definition of distant simultaneity, in section §2 of the 1905 relativity paper, Einstein presents in the following way the principal result of his kinematics: *relativity of simultaneity*.

Let there be a [theoretical] rigid rod at rest. Its length is $l$ as measured by a measuring rod which is also at rest. Einstein imagines the axis of the rod lying along the $X$ axis of the coordinate system at rest. The rod is then set into uniform parallel motion of translation (with velocity v) along the $X$-axis in the direction of increasing $x$. Einstein asks, what is the length of the *moving rod*? Einstein defines its length in two different ways by implementing his two heuristic principles:

a) First there is an observer who moves together with the rod to be measured and with the measuring rod. He measures the length of the rod directly by superposing the measuring rod, in just the same way as if the three were at rest.

b) "The observer in the rest system ascertains, as required by §1, using synchronous clocks at rest, at what points of the rest system the two ends of the rod to be measured are located at a definite time t. The distance between these two points, measured with the rod used before, but now at rest, is also a length, which may be designated the 'Length of the Rod' ".[21]

Einstein is guided by the principle of relativity and thus calls the length by operation a) – the length of the rod in the moving system. This length is equal to the length $l$ of the rod at rest. Einstein calls the length of the rod by operation b), "the length of the (moving) rod in the system at rest'.

Since, according to Einstein, one could give preference to either of the two observers, as the one was in motion and the other was at rest, this latter statement laid down the following definition: the length of a body in its rest frame is uncontracted.

Einstein was going to demonstrate the relativity of simultaneity by using both his principles: the relativity and the light postulate:[22]

Suppose the rod moves with a uniform velocity $v$ along the $x$ axis of the "rest system". At the two ends $A$ and $B$ of the moving rod, clocks are placed, are synchronous, however, with the clocks of the system at rest. With each clock there is a moving observer. Recall that observer $A$ can determine the time of events immediately near the clock $A$, and $B$ can do the same near clock $B$. Now we want the two clocks to be synchronized according to the procedure in §1 and so the observers exchange signals.

Imagine again "the observer in the system at rest" of b) who "ascertains, as required by §1" the kinematical length of the moving rod. The clocks of the observer in the system at rest are synchronous with the clocks of the observers *A* and *B*. He sees the two observers *A* and *B* exchanging light signals. In the moving system a ray of light is sent from *A* at time $t_A$, there it is reflected from *B* at time $t_B$, and arrives back at *A* at time $t'_A$. The observer in the system at rest uses the definition of distant simultaneity from §1 and finds:

$$t_B - t_A = [r_{AB} + v(t_B - t_A)]/c,$$

$$t'_A - t_B = [r_{AB} - v(t'_A - t_B)]/c,$$

where, $r_{AB}$ denotes the kinematical length of the moving rod, measured in the system at rest. Einstein wrote the equations in the following way: [23]

$$t_B - t_A = r_{AB}/(c - v),$$

$$t'_A - t_B = r_{AB}/(c + v).$$

This result shows that observers moving with the rod will thus find that their clocks are not synchronous, while observers in the system at rest will declare the clocks to be synchronous. "We thus see that we cannot ascribe any *absolute* meaning to the concept of simultaneity, but that two events, which are simultaneous as viewed from a system of coordinates, can no longer be considered simultaneous events when observed from a system which is moving relative to that system".[24]

## 3.2 Challenges on Einstein's Connection of Synchronization and Contraction

Einstein thought experiment shows the connection between the contraction of the moving rod as observed from reference frame K ("the system at rest") and the relativity of simultaneity. From this connection Vladimir Varičak suggested in 1911 the following interpretation to the contraction of lengths:

Einstein's contraction should be regarded as an apparent phenomenon, produced by the manner in which our clocks are synchronized and lengths are measured. It results from measuring length using synchronized clocks. This contraction is thus connected with the relativity of simultaneity. If a state is called real only when it can be determined in the same way in all reference frames, then the Lorentz contraction in Einstein's sense is only apparent. Because an observer in K will see the rod contracted, while an observer in k (co-moving with the rod) will see the rod without contraction. Thus is the Lorentz contraction observable?

Since the contraction is only observable from the standpoint of a non-co-moving observer, the contraction is a subjective phenomenon. It depends on an observer being present in order to measure length using synchronized clocks. On the other hand, Lorentz's contraction is an objective phenomenon, for it occurs whether an observer is present or not.[25]

On May 18 1911, Einstein submitted his paper, "On the Ehrenfest Paradox. Comment on V. Varičak's paper" to *Physikalische Zeitschrift*, in which he commented on Varičak's comments.[26]

Einstein thought that Varičak's comments should not go unanswered, because they may cause confusion. Varičak unjustifiably perceived a difference between the consequences of Lorentz's and Einstein's conception of the contraction with regard to the physical facts. The question of whether the Lorentz contraction exists or does not exist in reality is misleading, because it does not exist "in reality" insofar as it does not exist for an observer moving with the object. However, it does exist "in reality", in the sense that it could be detected by physical means by a non-co-moving observer.

Einstein ended his reply to Varičak's comments by suggesting the following thought experiment that does not require the use of synchronized clocks in order to verify the Lorentz contraction. Hence he showed that this contraction does not have its roots solely in the manner in which we synchronize our clocks: the relativity of simultaneity for which clocks are necessary for the observation of the Lorentz contraction in Einstein's 1905 paper, can be determined with the help of measuring rods only, without the use of clocks:[27]

Consider two rods A'B' and A"B" of the same proper length. The two rods move parallel to the x-axis relative to K with equal and opposite large constant velocity v: they past each other with A'B' moving in the positive, and A"B" in the negative direction of the x-axis. The endpoints A' and A" of the rods meet at a point A* on the x-axis of K, while the endpoints B' and B" meet at a point B* of K. The distance A*B* as measured by rods at rest in K will then be $\sqrt{1 - v^2/c^2}$ smaller than the proper length of either of the two rods.

Therefore, the Lorentz contraction is not a property of a single measuring rod taken by itself, but a reciprocal relation between two such rods A'B' and A"B" moving relatively to each other, and this relation is in principle observable.[28]

In 1916 Einstein used his definition of distant simultaneity with respect to a given frame of reference without any reference to clocks and suggested the famous train thought experiment to exemplify the relativity of simultaneity in his 1916/1917 popular book, *On the Special and the General Theory of Relativity*,[29]

We suppose a very long train traveling along the rails with constant velocity *v* relative to the earth. People traveling in the train consider events in reference to the train. Einstein considers two events, strokes of lightning *A* and *B*, and says that these are simultaneous with respect to the railway embankment, i.e., the earth frame. He asks, are these also simultaneous relatively to the train? He shows they are not. When we say that the lightning strokes *A* and *B* are simultaneous with respect to the embankment, we mean that the rays of light emitted at the places *A* and *B* – where the lightning occurs – meet each other at the midpoint *M* of the embankment at the same moment of time. But the events *A* and *B* correspond to positions *A* and *B* on the train.

Let *M′* be the midpoint of the length *AB* on the travelling train. When the flashes of lightning occur, as judged by an observer on the embankment, *M* and *M′* coincide, and then *M′* moves towards the right with velocity *v*. If an observer sitting in the position *M′* in the train did not possess this velocity, then he would remain at *M* and the light rays emitted by the two flashes of lightning *A* and *B* would reach him simultaneously, and they would meet him where he is seated at the same moment.

In reality the observer at *M′* is rushing towards the light signal coming from *B*, while he is riding on ahead of the beam of light coming from *A*. It means that the light signal from *A* have traversed a greater distance to arrive to the observer at *M′*. And the light beam from the place *B* traversed a lesser distance. Since the velocity of light is constant, the observer at *M′* sees the beam of light emitted from *B before* he sees that emitted from *A*. Einstein thus concludes: Observers who take the railway train as their reference frame must therefore arrive at the conclusion that the lightning flash B took place earlier than the lightning flash A.

## 4. Derivation of the Lorentz Transformation

### 4.1 1905: The Lorentz Transformations Derived by the Principle of Relativity and the Light Postulate

In section §3 Einstein derived the Lorentz transformations (the transformations of the coordinates and times) in the following way: He defined two systems: the first, a rest system K, and a second, system [k] moving with a constant velocity *v* in the direction of increasing *x* of the other system K. Einstein defined the two systems of coordinates and time of a specific event one with respect the system K and the other with respect to k. Einstein defined the two systems in the following manner: [30]

"Let there be in the space 'at rest' two systems of coordinates, i.e. two systems, each of three rigid material lines, perpendicular to one another, and originating from a point. Let the *X*-axes of the two systems coincide, and their *Y*- and *Z*-axes respectively be parallel. Each system shall be provided with a rigid measuring-rod and a number of clocks, and let both measuring-rods, and all the clocks of the two systems, be in all respects identical". Now impart to the origin of one of the two systems (*k*) a (constant) velocity *v* in the direction of the increasing *x* of the other system (*K*) at rest, and let this velocity be communicated to the axes of the coordinates, the relevant measuring-rod, and the clocks. To each time *t* of the system *K* at rest there then will correspond a definite position of the axes of the moving system, and from reasons of symmetry we are entitled to assume that the motion of *k* may be such that the axes of the moving system are at the time *t* (this "*t*" always denotes a time of the system at rest) parallel to the axes of the system at rest".

Einstein imagines that space is measured from the system $K$ at rest by means of a measuring-rod at rest, and from the moving system $k$ by means of the measuring-rod moving with it. The coordinates obtained by this way are: $x, y, z$ and $\xi, \eta, \zeta$, respectively.[31] The time is determined by means of light signals in the manner indicated in section §1 of the relativity paper: there are clocks at many points in system $K$ and observers exchanging light signals between these points at which the clocks are located – the time $t$ is determined for all these points in $K$ where there are clocks. The same method is used in system $k$, with clocks at rest relative to system $k$, and time $\tau$ is determined for $k$.[32] To any system of values $x, y, z, t$, which completely define the place and time of the event in the system $K$ at rest, there corresponds a system of values $\xi, \eta, \zeta, \tau$, determining that event relative to the system k. Einstein's task is to find the system of transformations equations – connecting these quantities.[33]

The derivation of the Lorentz transformation, or as Einstein calls them in 1905, the transformations of Coordinates and Time, in section §3 of the relativity paper – the equations connecting the two systems of quantities $x, y, z, t$ and $\xi, \eta, \zeta, \tau$ – is very cumbersome. It is based on synchronizing clocks, and expressing "in equations that $\tau$ is nothing but the collection of the readings of clocks at rest in system $k$, which have been synchronized according to the rule given in §1" (distant simultaneity).[34]

The derivation uses the two guiding principles: [35]

"[…] and applying the Principle of the Constancy of the Velocity of Light in the rest system", and "expressing in equations that light (as required by the Principle of the Constancy of the Velocity of Light in combination with the Principle of Relativity) is also propagated with velocity $c$ when measured in the moving system".

The final form of Einstein's transformation equations for the coordinates and time is:[36]

$$\tau = \beta \left( t - \frac{vx}{c^2} \right)$$

$$\xi = \beta (x - vt)$$

$$\eta = y$$

$$\zeta = z$$

$$\beta = \frac{1}{\sqrt{1 - v^2/c^2}}$$

## 4.2 Lorentz Transformations Derived without the Light Postulate

In 1910, the Russian physicist Vladimir (Waldemar) Ignatowski showed that the special theory of relativity does not require the light postulate.[37] In order to construct a theory of *relativity without light*, and to derive Lorentz-like transformations from the relativity principle alone, one needs to assume several additional hypotheses:[38]

1) The requirement of reciprocity: the velocity of K relative to K' is equal and opposite to that of K' relative to K; the contraction of lengths at rest in K' and observed in K is equal to that of lengths at rest in K and observed in K'.[39]

2) Isotropy and homogeneity of space and time.

As seen before, Einstein also needed additional hypotheses that were generally valid, transitivity and additivity.[40]

Ignatowski proved that Lorentz-like transformations follow from such assumptions, with one parameter α, which can be either finite or zero: [41]

$$x' = \frac{x - vt}{\sqrt{(1 - \alpha v^2)}}, \qquad t' = \frac{t - \alpha v x}{\sqrt{(1 - \alpha v^2)}}.$$

The parameter can be identified with the reciprocal of the maximum possible speed in inertial frames; and if we apply these Lorentz transformations to the Maxwell equations, then the parameter is the speed of light. Ignatowski thus claimed he proved that the second postulate of Einstein's theory was superfluous.

On the other hand, Einstein's conclusion, repeatedly demonstrated in the 1905 paper from a few points of view that, the speed of light is the upper limit on the velocities in inertial systems, is required in Ignatowski's formalism. However, one could claim that Ignatowski was forced to recourse to electrodynamics in order to include the speed of light, and thus indeed the second postulate of Einstein was necessary.

Torretti found inconsistencies and contradictions in Ignatowski's additional hypotheses 1) and 2), and that his derivation after all requires the additional hypothesis that of the invariant velocity.[42]

Since the value of the parameter can be zero or finite, the theory of relativity without light allows two possible transformations. For any finite value of the parameter, the group is isomorphic to the Lorentz group. As the parameter becomes α = 0, the group contracts to the Galilei group:

x' = x − vt, t' = t.

By putting $c = \infty$ in Einstein's Lorentz transformations they can equally well be derived from them, says Wolfgang Pauli in his 1921 Encyclopedia article on Relativity.[43]

Stachel said that Einstein, however, never gave up the light postulate, and accounts of the theory kinematics tended to follow Einstein's lead.[44]

## 5. Composition of Relative Velocities

### 5.1 1905: The Addition Theorem for Velocities

In section §5 of the 1905 relativity paper Einstein obtained the addition theorem for relative velocities. Along the *X* axis of the system *K*, a point is moving in the system *k* with a velocity *v* according to the equations:[45]

$\xi = w_\xi \tau$, $\eta = w_\eta \tau$, $\zeta = 0$, where, $w_\xi$ and $w_\eta$ denote constants.

Einstein required the motion of the point relative to the system *K*. With the help of the system of Lorentz transformations developed in section §3, Einstein obtained for the equations of motion of the point:

$$x = \frac{t(w_\xi + v)}{(1 + vw_\xi/c^2)}, y = \frac{w_\eta t\sqrt{1 - v^2/c^2}}{(1 + vw_\xi/c^2)}, z = 0$$

and:

$U^2 = (dx/dt)^2 + (dy/dt)^2$, $w^2 = w^2_\xi + w^2_\eta$

Einstein finally obtained U if *w* has the direction of the *X*-axis:

$U = (v + w)/(1 + vw/c^2)$.[46]

After presenting the addition theorem of velocities in one direction, not in the general case, he introduced a third system of coordinates *k'*, moving parallel to *k*. He wrote: "such parallel transformations – necessarily – form a group".

Einstein arrived at this conclusion by obtaining the formula for *U* if w and v have the same direction; by composition of two transformations from §3: *k'* that moves parallel to *k* with velocity *w* along the *Ξ*–axis of *k*, and *k* moving with velocity *v* along the *X*-axis of *K*, while *v* is replaced by: *(v + w)/(1 + vw/c^2)*.[47]

Einstein concluded, "It follows from this equation that from a composition of two velocities which are smaller than *c*, there always results a velocity smaller than *c*. [U is smaller than c…] It also follows that the velocity of light *c* cannot be changed by composition with a velocity less than that of light" [U = c always for anything moving with speed c]".[48]

### 5.2 Space of Relative Velocities

Stachel says that it is better calling the chapter title of section §5 of the relativity paper, "composition of relative velocities", because it is a group operation. However, Einstein used the classical Newtonian word "addition". The formulae expressing the composition of relative velocities are quite complicated algebraically, especially if the two velocities do not lie in the same direction. [49]

In 1909 Sommerfeld found a simple interpretation to the law of composition of relative velocities: The two velocities that are composed and their resultant form the sides of a spherical triangle of an imaginary radius $ic$. Sommerfeld based his assumption on Minkowski's studies, according to which the Lorentz-Einstein transformations are "space-time rotation" in a four-dimensional space xyzl, where l = ict, also means length (the light path multiplied by the imaginary unit i).[50]

Sommerfeld wrote the transformation equations:

$$x' = x \, cos\varphi + l \, sin\varphi, y' = y$$

$$l' = -x \, sin\varphi + l \, cos\varphi, z' = z.$$

There exists a relation between the imaginary rotation angle φ and the velocity ratio β:

$$tang\varphi = i\beta, cos\varphi = \frac{1}{\sqrt{1-\beta^2}}, sin\varphi = \frac{i\beta}{\sqrt{1-\beta^2}}$$

Sommerfeld then showed that if we compose two rotations of equal plane, then the rotation angles sum up. The same holds for two translations of equal direction (two space-time rotations in the same plane).

Therefore if $\varphi_1$ and $\varphi_2$ are the imaginary rotation angles, and φ the resulting angle from the composition, and $\beta_1, \beta_2, \beta; v_1, v_2, v$ are the corresponding velocity ratios and velocities, then,

$\varphi_2 + \varphi_1 = \varphi$

and using the imaginary form of Minkowski space Einstein's composition law of relative velocities can be formulated in terms of triangular addition on the surface of a sphere of imaginary radius,

$$\beta = \frac{1}{i} tang(\varphi_1 + \varphi_2) = \frac{1}{i} \frac{tang\,\varphi_1 + tang\,\varphi_2}{1 - tang\,\varphi_1 tang\,\varphi_2} = \frac{\beta_1 + \beta_2}{1 + \beta_1\beta_2}$$

and

$$v = \frac{v_1 + v_2}{1 + v_1 v_2/c^2}.$$

In 1910 in his paper, "Application of Lobachevskian geometry in the Theory of Relativity" Varičak applied hyperbolic geometry to the special theory of relativity,[51]

"For the composition of velocities in the theory of relativity, the formulas of spherical geometry with imaginary sides are valid, as it has been recently shown by Sommerfeld in this Journal. Now, the non-Euclidean Geometry of Lobachevsky and Bolyai is the imaginary counter-image of the spherical geometry, and it is easily seen that an interesting field of application offers itself for the hyperbolic geometry.

As relative motion of reference frames with superluminal speed does not occur, we can always put,

$\frac{v}{c} = tanh u$

The factor $\frac{1}{\sqrt{1 - \frac{v^2}{c^2}}}$ that plays an important role in the Lorentz-Einstein transformation equations and the formulas derived from them, goes over into cosh u if we in addition put i=ct, then the transformation equations read,

$l' = -x\, sinh\, u + l\, cosh\, u$

$x' = x\, cosh\, u - l\, sinh\, u$

$y' = y, z' = z.$

[…]

The hyperbolas that are invariant with respect to these transformations are,

$\omega(l, x) \equiv l^2 - x^2 + \text{const} = 0.$"

In his 1912 paper, "On the Non-Euclidean Interpretation of the Theory of Relativity", Varičak reinterpreted Sommerfeld's 1909 result (i.e., formulating the composition of relative velocities in terms of triangular addition on the surface of a sphere of imaginary radius) by applying his above 1910 study. Varičak connected between the composition of relative velocities and non-Euclidean hyperbolic space; this is a geometric approach to solving problems involving the composition of relative velocities.[52]

Varičak then defined the quantity,

$u = \tanh^{-1} \frac{v}{c}.$

For tanh u = 1 then v =c.

He then wrote,

$\frac{v_1}{c} = \tan h\, u_1, \frac{v_2}{c} = \tan h\, u_2.$

From examining triangles he arrived at:

$$cosh u = cosh u_1 cosh u_2 + sinh u_1 sinh u_2 cosh \alpha$$

and,

$$cosh u = \frac{1}{\sqrt{1 - \left(1 - \frac{v^2}{c^2}\right)}}, sinh u = \frac{\frac{v}{c}}{\sqrt{1 - \left(1 - \frac{v^2}{c^2}\right)}}$$

And from this Varičak eventually arrived at Einstein's general law for composition of relative velocities,

$$v = \frac{\sqrt{v_1^2 + v_2^2 + 2v_1 v_2 cos\alpha - \left(\frac{v_1 v_2 sin\alpha}{c}\right)^2}}{1 + \frac{v_1 v_2 cos\alpha}{c^2}}$$

If one neglects the last term in the numerator and denominator, and c the parameter of the Lobachevskian space is infinitely small, then it tends to a Euclidean space. In this case $\alpha = 0$, between the velocities $v_1$ and $v_2$, which lie in the same direction. And thus we obtain,

$$\tanh u = \frac{\tanh u_1 + \tanh u_2}{1 + \tanh u_1 \tanh u_2}.$$

And this is equivalent to,

$$v = \frac{v_1 + v_2}{1 + \frac{v_1 v_2}{c^2}}.$$

Alfred Robb reinterpreted Varičak's above procedure and formulas in terms of the "rapidity",[53]

"We propose now to define what we shall call the *rapidity* of a particle with respect to a system of permanent configuration.

*If v be the absolute velocity of the particle with respect to the system, then the inverse hyperbolic tangent of v will be spoken of as the rapidity.*

Thus if ω be the rapidity,

v = tanh ω." (c = 1).

The rapidity is Varičak's $u = \tanh^{-1} \frac{v}{c}$. (Robb's ω = Varičak's u).

After writing formulas for the composition of relative velocities in terms of hyperbolic trigonometric identities and the rapidity, Robb concludes,[54]

"*Thus instead of a Euclidean triangle of velocities, we get a Lobatchefckij triangle of rapidities. For small rapidities, however, we may identify rapidity and velocity, and the Lobatchefckij triangle may be treated as a Euclidean one. It is also seen that rapidities in the same straight line are additive*".

Therefore, relative velocities and composition of relative velocities correspond to certain properties of hyperbolic triangles in velocity space. Newtonian-Galilei velocity space is then flat, leading to the Galilean law of vector addition of relative velocities.

Robb then concluded, [55]

"The formulae which we have obtained agree with those of Einstein if we take the "*velocity of light*" as unity and express the results in terms of velocities of rapidities".

*I wish to thank Prof. John Stachel from the Center for Einstein Studies in Boston University for sitting with me for many hours discussing special relativity and its history. Almost every day, John came with notes on my draft manuscript, directed me to books in his Einstein collection, and gave me copies of his papers on Einstein, which I read with great interest*

[1] Einstein, Albert, "Zur Elektrodynamik bewegter Körper, *Annalen der Physik* 17, 1, 1905, pp. 891-921, p. 892.

[2] Einstein, 1905, p. 892.

[3] Stachel, John, "Special Relativity From Measuring Rods", in Cohen, R. S., and Laudan, L. (eds), *Physics, philosophy and Psychoanalysis*, 1983, pp. 255-272, p. 259.

[4] Born, Max, "Die Theorie des starren Elektrons in der Kinematik des Relativitäsprinzips", *Annalen der Physik* 30, 1909, pp. 1-56; p. 2.

[5] Sommerfeld, Arnod, "Diskussion, *Physikalische Zeitschrift* 10, 1909, pp. 826-829.

[6] Born, Max, "Über die Definition des starren Körpers in der Kinematik des Relativitäsprinzips", *Physikalische Zeitschrift* 11, 1910, pp. 233-234.

[7] Laue Max, "Zur Diskussion über den starren Körporen in der Relativitätstheorie", *Physikalische Zeitschrift* 12, 1911, pp. 85-87; p. 86.

[8] Laue, 1911, p. 86.

[9] Laue, 1911, p. 87.

[10] Einstein, Albert, *Relativity, The Special and the General Theory*, Fifteenth Edition, trans. Robert W. Lawson, 1952, New-York: Crown Publishers, Inc, pp. 142-143.

[11] Einstein, 1905, pp. 892-893.

[12] Einstein, 1905, p. 893.

[13] Stachel, John, "Einstein's Clocks, Poincare's Maps/Empires of Time", *Studies in History and Philosophy of Science* B 36 (1), pp. 202-210; pp. 1-15. And Martínez says that, "Einstein himself left no written statement even implying that he became increasingly interested in time because of any timing technologies at the patent office". Martínez, Alberto, "Material History and Imaginary Clocks: Poincaré, Einstein, and Galison on Simultaneity", *Physics in Perspective* 6, 2004, pp. 224-240, p. 226.

[14] Einstein, 1905, p. 893.

[15] Einstein, 1905, pp. 893-894.

[16] Einstein, 1905, p. 894.

[17] Einstein, 1905, p. 894.

[18] Einstein, 1905, p. 894.

[19] Einstein, Albert, *Uber die Spezielle und die Allgemeine Relativitätstheorie, Gemeinverständlich* ,1920, Braunschweig: Vieweg Shohn, pp. 14-15; Stachel, 2005, pp. 7-8.

[20] Einstein, 1905, p. 895.

[21] Einstein, 1905, pp. 895-896.

[22] Einstein, 1905, p. 896.

[23] Einstein, 1905, pp. 896-897.

[24] Einstein, 1905, pp. 897.

[25] Varičak, Vladimir, "Zum Ehrenfestschen Paradoxon", *Physikalische Zeitschrift* 12, 1911, p. 169, p. 169.

[26] Einstein, Albert, "Zum Ehrenfestschen Paradoxon. Bemerkung zu V. Varičak's Aufsatz",. *Physikalische Zeitschrift* 12, 1911, pp. 509-510; pp. 509-510.

[27] "Es seinen zwei (ruhend verglichen) gleichlange stäbe A'B' und A"B", welche längs der X-achs eines beschleunigungsfreien Koordinaten-system in der X-Achse paralleler. gleichsinniger Orientierung gleiten konnen. A'B' und A"B" sollen aneinander vorbeigleiten, wobei A'b' im Sinne der positiven, A"B" im Sinne der negativen X-Achse mit beliebig großer konstanter Geschwindigkeit bewegt sei. Dabei begegnen sich die endpunkte A' und A" in einem Punkte A* die Endpunkte B' und B" in einem Punkte B* der X-Achse. Die Entfernung A*B* ist dann nach der Relativitätstheorie kleiner als die Länge eines jeden der Stäbe A'B' und A"B", was mit einem der Stäbe konstatiert werden kann, indem derselbe im Zustand der Ruhe an der Strecke A*B* angelegt wird". Einstein, 1911, p. 510.

[28] Pauli, Wolfgang (1921/1958), "Relativitätstheorie" in *Encyklopädie der Mathematischen wissenschaften*, Vol. 5, Part 2, 1921, Leipzig: Teubner, 2000, p. 27 (p. 557); Pauli, Wolfgang, English translation, *Theory of Relativity*, 1958, Oxford and New York: Pergamon, pp. 12-13;

[29] Einstein, 1917, pp. 16-17.

[30] Einstein, 1905, p. 897.

[31] Einstein, 1905, pp. 897-898.

[32] Einstein, 1905, p. 898.

[33] Einstein, 1905, p. 898.

[34] Einstein, 1905, p. 898.

[35] Einstein, 1905, p. 899.

[36] Einstein, 1905, p. 902.

[37] Stachel, John, "History of relativity," in Brown, Laurie M., Pais, Abraham, and Sir Pippard, Brian (eds.), *Twentieth century physics*, 1995, New York: American Institute of Physics Press, pp. 249-356; p 277.

[38] Ignatowski, Vladimir, "Einige Allgemine Bemerkungen über das Relativitätsprinzip, *Physikalische Zeitschrift* 11, pp. 972-976; "Eine Bemerkung zu meiner Arbeit 'Einige Allgemeine Bemerkungen über das Relativitätsprinzip'", *Physikalische Zeitschrift* 12, p. 779.

[39] Pauli, 2000, pp. 25-26 (pp. 555-556); Pauli, 1921/1958, p. 11.

[40] Einstein, 1905, p. 894.

[41] Pauli, 2000, p. 26 (p. 556); Pauli, 1921/1958, p. 11.

[42] Torretti, Roberto, *Relativity and Geometry*, 1983/1996, Ney-York: Dover, pp. 298-299.

[43] Pauli, 2000, p. 26 (p. 556); Pauli, 1921/1958, p. 11.

[44] Stachel, 1995, pp. 277-278.

[45] Einstein, 1905, p. 905.

[46] Einstein, 1905, pp. 905-906.

[47] Einstein, 1905, pp. 906-907.

[48] Einstein, 1905, p. 906.

[49] Stachel, 1995, p. 278.

[50] Sommerfeld, Arnold, "über die Zusammensetung der Geschwindigkeiten in der Relativtheorie", *Berichte der Deutschen Physikalischen Gesellschaft*, pp. 577-582; pp. 577-578.

[51] Varičak, V, "Anwendung Lobatschefskijschen Geometrie in der Relativtheorie", *Physikalische Zeitschrift* 11, 1910, pp. 93-96; p. 93.

[52] Varičak, V, "Über die nichteuklidische Interpretation der Relativtheorie", *Jahresbericht der Deutschen Mathematiker-Vereinigung* 21, 1912, pp. 103-127; pp. 106-109.

[53] Robb, Alfred, *Optical Geometry of Motion; a New View of the Theory of Relativity*, 1911, London: Cambridge, W. Heffer and Sons LTD, pp. 8-9.

[54] Robb, 1911, p. 29.

[55] Robb, 1911, p. 29.